\documentclass[preprint, aps, nofootinbib, prb]{revtex4-1}
\usepackage{amsmath} \usepackage{graphicx} \usepackage{dcolumn}
\usepackage{bm} 
\usepackage{amssymb}
\usepackage{pstricks}
\usepackage{siunitx} 
\usepackage{braket}
\usepackage{subfigure}

\newrgbcolor{dred}{.67 0 0}

\makeatletter

\renewenvironment{widetext@grid}{%
  \par\ignorespaces
  \setbox\widetext@top\vbox{%
   \vskip15\p@
   \hb@xt@\hsize{%
    \leaders\hrule\hfil
    \vrule\@height6\p@
   }%
   \vskip6\p@
  }%
  \setbox\widetext@bot\hb@xt@\hsize{%
    \vrule\@depth6\p@
    \leaders\hrule\hfil
  }%
  \onecolumngrid
  \let\set@footnotewidth\set@footnotewidth@ii
}{%
  \par
  \twocolumngrid\global\@ignoretrue
  \@endpetrue
}%

\makeatother

\begin{document}

\newlength{\figurewidth}
\setlength{\figurewidth}{0.6 \columnwidth}

\newcommand{\prtl}{\partial}
\newcommand{\la}{\left\langle}
\newcommand{\ra}{\right\rangle}
\newcommand{\dla}{\la \! \! \! \la}
\newcommand{\dra}{\ra \! \! \! \ra}
\newcommand{\we}{\widetilde}
\newcommand{\smfp}{{\mbox{\scriptsize mfp}}}
\newcommand{\smp}{{\mbox{\scriptsize mp}}}
\newcommand{\sph}{{\mbox{\scriptsize ph}}}
\newcommand{\sinhom}{{\mbox{\scriptsize inhom}}}
\newcommand{\sneigh}{{\mbox{\scriptsize neigh}}}
\newcommand{\srlxn}{{\mbox{\scriptsize rlxn}}}
\newcommand{\svibr}{{\mbox{\scriptsize vibr}}}
\newcommand{\smicro}{{\mbox{\scriptsize micro}}}
\newcommand{\scoll}{{\mbox{\scriptsize coll}}}
\newcommand{\sattr}{{\mbox{\scriptsize attr}}}
\newcommand{\sth}{{\mbox{\scriptsize th}}}
\newcommand{\sauto}{{\mbox{\scriptsize auto}}}
\newcommand{\seq}{{\mbox{\scriptsize eq}}}
\newcommand{\teq}{{\mbox{\tiny eq}}}
\newcommand{\sinn}{{\mbox{\scriptsize in}}}
\newcommand{\suni}{{\mbox{\scriptsize uni}}}
\newcommand{\tin}{{\mbox{\tiny (in)}}}
\newcommand{\tout}{{\mbox{\tiny (out)}}}
\newcommand{\scr}{{\mbox{\scriptsize cr}}}
\newcommand{\tstring}{{\mbox{\tiny string}}}
\newcommand{\sperc}{{\mbox{\scriptsize perc}}}
\newcommand{\tperc}{{\mbox{\tiny perc}}}
\newcommand{\sstring}{{\mbox{\scriptsize string}}}
\newcommand{\stheor}{{\mbox{\scriptsize theor}}}
\newcommand{\sGS}{{\mbox{\scriptsize GS}}}
\newcommand{\sBP}{{\mbox{\scriptsize BP}}}
\newcommand{\sNMT}{{\mbox{\scriptsize NMT}}}
\newcommand{\sbulk}{{\mbox{\scriptsize bulk}}}
\newcommand{\tbulk}{{\mbox{\tiny bulk}}}
\newcommand{\sXtal}{{\mbox{\scriptsize Xtal}}}
\newcommand{\sliq}{{\text{\tiny liq}}}

\newcommand{\smin}{\text{min}}
\newcommand{\smax}{\text{max}}

\newcommand{\saX}{\text{\tiny aX}}
\newcommand{\slaX}{\text{l,{\tiny aX}}}

\newcommand{\svap}{{\mbox{\scriptsize vap}}}
\newcommand{\sjam}{J}
\newcommand{\Tm}{T_m}
\newcommand{\sTS}{{\mbox{\scriptsize TS}}}
\newcommand{\sDW}{{\mbox{\tiny DW}}}
\newcommand{\cN}{{\cal N}}
\newcommand{\cB}{{\cal B}}
\newcommand{\br}{{\bm r}}
\newcommand{\bR}{{\bm R}}
\newcommand{\phloc}{\phi^\text{(loc)}}
\newcommand{\ptloc}{\we \phi^\text{(loc)}}

\newcommand{\be}{\bm e}
\newcommand{\cH}{{\cal H}}
\newcommand{\cHlt}{\cH_{\mbox{\scriptsize lat}}}
\newcommand{\sthermo}{{\mbox{\scriptsize thermo}}}

\newcommand{\bu}{\bm u}
\newcommand{\bk}{\bm k}
\newcommand{\bq}{{\bm q}}
\newcommand{\bX}{\bm X}
\newcommand{\bY}{\bm Y}
\newcommand{\bA}{\bm A}
\newcommand{\bb}{\bm b}

\newcommand{\lintf}{l_\text{intf}}

\newcommand{\DV}{\delta V_{12}}
\newcommand{\sout}{{\mbox{\scriptsize out}}}
\newcommand{\dv}{\Delta v_{1 \infty}}
\newcommand{\dvin}{\Delta v_{2 \infty}}

\newcommand*\xbar[1]{%
  \hbox{%
    \vbox{%
      \hrule height 0.5pt 
      \kern0.5ex
      \hbox{%
        \kern-0.1em
        \ensuremath{#1}%
        \kern-0.1em
      }%
    }%
  }%
}

\newcommand{\cV}{{\cal V}}

\def\Xint#1{\mathchoice
   {\XXint\displaystyle\textstyle{#1}}%
   {\XXint\textstyle\scriptstyle{#1}}%
   {\XXint\scriptstyle\scriptscriptstyle{#1}}%
   {\XXint\scriptscriptstyle\scriptscriptstyle{#1}}%
   \!\int}
\def\XXint#1#2#3{{\setbox0=\hbox{$#1{#2#3}{\int}$}
     \vcenter{\hbox{$#2#3$}}\kern-.5\wd0}}
\def\ddashint{\Xint=}
\def\dashint{\Xint-}
\title{Cavitation in Electron Fluids and \\ the Puzzles of
  Photoemission Spectra in Alkali Metals}

\author{Roman Dmitriev} \affiliation{Department of Chemistry, University
  of Houston, Houston, TX 77204-5003}

\author{Jenny Green} \affiliation{St. John's School, Houston, TX
  77019} \altaffiliation[Current address: ] {Department of Electrical
  \& Computer Engineering, Duke University, Durham, NC 27708}

\author{Vassiliy Lubchenko} \email{vas@uh.edu} \affiliation{Department
  of Chemistry, University of Houston, Houston, TX 77204-5003}
\affiliation{Department of Physics, University of Houston, Houston, TX
  77204-5005} \affiliation{Texas Center for Superconductivity,
  University of Houston, Houston, TX 77204-5002}

\date{\today}

\begin{abstract}

  Angle-resolved photoemission spectra of alkali metals exhibit a
  puzzling, non-dispersing peak in the apparent density of states near
  the Fermi energy.  We argue that the holes left behind a significant
  fraction of photoejected electrons are not wavepacket-like objects
  used to describe excitations of an equilibrium Fermi liquid but,
  instead, are relatively localized entities resulting from a
  photon-induced cavitation in the electron fluid. At the same time,
  these special localized holes can be thought of as vacancies in a
  transient Wigner solid. The corresponding contribution to the
  photoemission current is non-dispersive  and is tied to the
    Fermi level; it exhibits certain similarities to photoemission
  from localized core orbitals such as the presence of recoil
  currents. { Calculated spectra are consistent with experiment.
  } We briefly discuss the present findings in the context of quantum
  measurement.

\end{abstract}

\maketitle

Angle-resolved spectra of sodium and potassium exhibit a distinct,
peculiar peak near the Fermi energy, within a substantial range of
photon energies.~\cite{PlumFirst, LyoKexperiment,
  1402-4896-1987-T17-021} This peak does not move with the photon's
energy and is narrower than the momentum-conserving, dispersing peak.
The intensity of the non-dispersing peak is substantial even when the
initial state of the electron for a {\em vertical} transition would be
above the Fermi energy.  Furthermore in potassium, it is the
dispersing peak that is often hard to resolve,~\cite{LyoKexperiment}
while the anomalous peak is clearly visible.  The Fermi surfaces in Na
and K are particularly simple---nearly spherical in fact---and fully
contained within the first Brillouin zone;~\cite{Ashcroft} thus no
sharp features in the density of states are expected. Surface states
do not seem to be at play either, since the resulting peaks, if any,
would not be strictly tied to the Fermi energy.

Mahan and coworkers~\cite{PhysRevLett.57.1076, PhysRevB.36.4499}
argued that the peculiar photoemission peak is a wing of the
dispersing peak when the latter is centered at electron energies above
the Fermi surface; the broadening is due to interactions and their
interplay with the free surface.  Detailed
estimates~\cite{PhysRevLett.57.1076, PhysRevB.36.4499} yield
photoemission spectra that are qualitatively similar to some of those
observed in sodium for photon energies corresponding to the $(1, 1, 0)
\to (3, 3, 0)$ transitions, but lack the anomalous ``balcony peaks''
seen in the adjacent range of photon energies that correspond to the
$(1, 1, 0) \to (-4, -4, 0)$ transitions.~\cite{PhysRevLett.58.959} The
situation with potassium is even less
conclusive.~\cite{PhysRevB.50.5004}

Overhauser~\cite{PhysRevLett.55.1916} proposed, alternatively, that
the anomalous peak results from a static charge density wave
(CDW).~\cite{OverhauserAP} Sodium does become close-packed
at sufficiently low temperatures thus implying, potentially, a
structural instability. Still, experimental
studies~\cite{PhysRevB.13.1017} decisively rule out the presence of a
static CDW, consistent with recent studies~\cite{PhysRevB.101.220103}
according to which the Fermi surface of sodium is relatively
insensitive to the detailed structure of the crystal. { A
  detailed review of previous work can be found in
  Ref.~\cite{1402-4896-1987-T17-021} }

{ Here we argue that the puzzling photoemission peak is caused by
  non-adiabatic effects that are not amenable to perturbative
  expansions around the equilibrium state of the electron assembly.}
The frequency $\omega_\text{ph}$ of the incoming photon is much
greater than the typical rates of electronic motions,
$\omega_\text{ph} \gg v_F/a$, where $v_F$ and $a$ are the Fermi
velocity and lattice spacing, respectively. Thus one expects a
response similar to giant resonances seen in nuclear
spectra,~\cite{AndersonBasicNotions} though spanning a relatively
narrow spectral range because the plasma oscillations are in their
ground state at the energies in question.  To quantify the
photocurrent one must compute the one-particle density function
$\rho(\br_1, \br_2)$ of the electron. The momentum-like argument $\bk$
of the Wigner transform of the latter density matrix, $\rho(\br, \bk)
= \int d^3 (\br_2 - \br_1) \rho(\br_1, \br_2) e^{-i \bk (\br_2 -
  \br_1)}$, essentially corresponds to the momentum $\bk$ of the
electronic wavepacket, while the dependence of the latter Wigner
transform on the center-of-mass variable $\br=(\br_1+\br_2)/2$
reflects the spatial variation of the corresponding charge
density. Landau's Fermi-liquid theory corresponds to the limit of this
spatial variation being very slow and describes the quasi-equilibrium
response of the electron fluid.~\cite{Goldstone1959CollectiveEO}
Conversely, high-frequency motions of the electron fluid, $\omega >
v_F/a$, are heavily hybridized with the plasmons via Landau
damping,~\cite{PinesEXS, nozieres1999theory} whereby the charge
density varies on length scales comparable to the lattice
spacing. This, then, suggests a possibility that in addition to
extended electron wavepackets characteristic of the equilibrium,
Fermi-liquid behavior, the fast photons can also knock out individual
electrons in the form of localized entities.

In fact, just this latter possibility is the only one that could be
realized classically. For concreteness, we consider a setup in which a
compact region of a Newtonian fluid changes its velocity
instantaneously from zero to $v_0$, { as it would in response to
  a sudden perturbation applied to the region}. Specifically for a
spherically-shaped region of radius $R$, the reactive force of the
surrounding fluid depends on time in the following manner, per the
solved problem 24.9 from Ref.~\cite{LLhydro}:
\begin{equation} \label{hydro} F(t) = 6 \pi \eta R v_0 \left[1 + R \sqrt{\rho/t
      \pi \eta} \right] + \frac{2 \pi}{3} \rho R^3 v_0 \delta(t)
\end{equation}
where $\eta$ and $\rho$ are the viscosity and density, respectively,
of the fluid. The intensity of dissipation is given by $v_0 F(t)$. Of
interest here is the non-adiabatic contribution $\propto t^{-1/2}$  to
the viscous part of the response, which represents a characteristic
hydrodynamic tail and, tellingly, scales with the area of the
sphere. The corresponding loss spectrum amounts to an inverse-square
root peak $\omega^{-1/2}$, which diverges at low energies. The latter
low energies would correspond to the vicinity of the Fermi energy in
an electron fluid.

{ This notion prompts us to inquire whether a classical-like
  photoemission from localized electronic states can occur---as a bulk
  phenomenon---in quantum fluids made of electrons. It would suffice
  for such localization, if any, to be only transient,
  because of the short duration of photoemission events. Localization
  of particles in the bulk simply means the particles have formed a
  solid. A solid is a state of broken translational symmetry in which
  the particles are each assigned to specific sites in space; the
  particles perform vibrational motions around their respective
  sites.~\cite{L_AP} The notion of a solid-like component to the
  wavefunction of an electron fluid may seem surprising, at first. We
  recall however that the electric current, if any, is exclusively due
  to the electrons' ability to tunnel through classically forbidden,
  inter-nuclear regions. One may associate the time spent in
  classically-forbidded regions with the liquid component of the
  wavefunction. The remaining time electrons perform bound motions
  within the classically allowed regions, each region assigned to a
  corner of a lattice. These motions correspond to a solid-like
  component of the overall wave function and lower translational
  symmetry, even if transiently.

Let us construct the solid-like component for a monovalent solid,
which houses one electron per site. Begin with a Wigner solid of the
jellium, whereby the electrons are sufficiently far apart and the
positive charge is uniformly distributed.~\cite{AndersonBasicNotions}
Imagine a process where we uniformly compress the system. To
compensate for the concomitant increase in the kinetic energy of the
electrons, we redistribute the positive charge so as to create a local
excess of positive charge at the lattice sites of the original Wigner
solid; the sites will become the actual atomic nuclei at the end of
the compression process. (The number of sites remains constant during
the process. The pertinent Wigner solid does not have to be strictly
periodic,~\cite{SchmalianWolynes2000} thus allowing for vibrational
displacements of the nuclei and a variety of cell shapes.)  Since the
barriers in the crystal field that separate distinct lattice sites are
finite, a fluid component to the electronic wavefunction will appear
eventually. Because the number of electrons per site remains constant,
the ``compression construct'' represents a continuous process; thus
the solid-like and the liquid-like components of the electron assembly
coexist, when both are present. The two phases remain in mutual
equilibrium, while their respective mole fractions depend on the
extent of the compression. This effective coexistence of two distinct
phase behaviors in the very same region of space is analogous to what
happens during the crossover to activated transport in
liquids,~\cite{L_AP, LW_ARPC} when metastable structures begin to
form. Translational symmetry is broken on times scales shorter than
the lifetimes of the metastable structures but is restored on longer
times.  Note a co-existence of liquid and solid behaviors,
respectively, has been reported for Hartree-Fock solutions in
jellium.\cite{doi:10.1002/ctpp.201700139} }

Within a single-electron picture, the localized states can be thought
of as bound states individual electrons transiently create for each
other on short times. (This is in addition to the potentials due to
the ionic cores, of course). The corresponding energy levels are,
however, not well-defined because the electrons are not static. The
resulting line broadening is analogous to what happens during spectral
diffusion,~\cite{KlauderAnderson, LS_EnMass} since the leading
contribution of local charge fluctuation to the shift of on-site
energies is dipole-dipole, owing to charge conservation. An electron
moves from site to site at rate $v_F/a$, while inducing a local dipole
moment change $e a$. These effective dipoles are uniformly distributed
at concentration $n \sim 1/a^3$. Contributions of individual dipoles
to the overall spectral shift are roughly $\sim (a e)^2/r^3 \equiv
A/r^3$ each fluctuating at rate $\gamma \sim v_F/a$.  The width of the
corresponding spectral line increases with time~\cite{KlauderAnderson,
  LS_EnMass} at the rate $\sim n A \gamma = (e^2/a) (v_F/a) \simeq E_F
(v_F/a)$. The broadening on the time scale $\pi/\omega_\text{ph}$ of a
photoemission event is, then, roughly $E_F (v_F/a \omega_\text{ph})
\sim 10^{-1} E_F$, consistent with experiment. { The solid-like response
will be progressively diminished for slower experimental probes, the
overall response ultimately approaching that of a Fermi liquid.

The liquid-like and solid-like contributions to the wavefunction
correspond to two distinct, non-overlapping components of the overall
wavefunction characterized by pronounced localization in the momentum
and direct space, respectively. Indeed, already the ground state of an
electron fluid in the presence of a scattering potential is orthogonal
to the ground state in the absence of the
potential.~\cite{AndersonInfraredCatastrophe} Very generally, a solid
must be separated by a discontinuous transition from the fluid
state.~\cite{L_AP, LandauPT1, Brazovskii1975}} Consequently, the two
phases occupy disconnected portions of the phase space. Because of
this lack of overlap between the liquid-like and solid-like
contributions to the overall wavefunction, we may present the total
intensity of the photocurrent as a weighted sum of the respective
intensities of those two contributions:
\begin{equation} \label{Itotal} I(E)=x_\text{liq} I_\text{liq}(E) +
  x_\text{sol} I_\text{sol}(E),
\end{equation}
where $E$ is the energy of the detected electron and $x_\text{liq} +
x_\text{sol} = 1$, by construction.

To estimate the solid-like contribution $I_\text{sol}(E)$ to the
photocurrent, we first note that in the spectral range in
question~\cite{PlumFirst, LyoKexperiment, 1402-4896-1987-T17-021} no
plasmons are produced. { Indeed, the plasmon frequency in Na, 5.7
  eV,~\cite{doi:10.1021/jp810808h} is significantly greater than the
  Fermi energy, 2.8 eV.~\cite{PlumFirst}} In other words, our
transient electron solid recoils as a whole.  We will approximate this
solid as harmonic. The coordinates of any lattice fragments thus obey
the Gaussian distribution;~\cite{RL_Tcr} denote the corresponding
variance with $\delta r^2$. The probability for the solid to recoil as
a whole, after the fragment absorbs or emits momentum $q$, is given by
$e^{-q^2 (\delta r)^2}$, a notion used in M\"ossbauer
spectroscopy.~\cite{feynman1998statistical} Conversely, the expression
$e^{-q^2 (\delta r)^2}$ can be viewed, up to a multiplicative factor,
as the probability distribution for a local harmonic degree of freedom
$\delta r$ that is compatible with a zero-phonon recoil of the lattice
at momentum $q$. The corresponding ground state wave function is
$\psi_q(\br) = (q^2/\pi)^{3/4} \, e^{-q^2 r^2/2}$, where $q$
represents a parameter.  We thus estimate the photocurrent
$I_\text{sol}(E)$, due to localized initial states, by first
evaluating the current using the function $\psi_q(\br)$ as the initial
state, for a given value of $q$, and then averaging over a pertinent
distribution of $q$. The set of recoil values $q$, due to emitting a
localized electron, should be consistent with the rate of spatial
variation of the valence electrons. At values $k_f$ of the momentum of
the outgoing electron pertinent to Plummer et al.'s experiment, $k_f
\sim 3 k_F$, the valence wave function can be largely approximated by
the frontier atomic orbital ${\psi}_\text{fr}$ on an individual
center. (For Na, this would be the 3s orbital.) Thus we use the
magnitude squared of the normalized Fourier transform
$\left|\widetilde {\psi}_\text{fr} (q)\right|^2$, times $4 \pi q^2$,
as the probability distribution for the parameter $q$.

The (zero-plasmon) recoil due to the transient-solid component of the
electron assembly causes a negligibly small shift of the photoemission
spectrum, as does the recoil due to the pertinent nuclei.  If it were
not for the recoil due to the electron fluid, localized electrons
would be all extracted near the Fermi energy, the latter nominally
corresponding to a quiescent fluid devoid of currents, consistent with
the classical limit considered above. This notion can be also
formulated quantum-mechanically, in an effective single-electron
picture: Single-particle states for the extended and localized states
tend to mutually repel,~\cite{GHL, Mott1993} while the extended states
form a continuous band. Our effective localized states---which note
are not tied to lone pairs, impurity levels, or surface states
etc.---are thus ``pushed'' outside of the continuous band. At the same
time, there should be no gap between the delocalized and localized
states either, because the two sets of states correspond,
respectively, to a liquid and solid that co-exist, as already
mentioned. Consequently, the chemical potentials of the phases are
mutually equal, which pegs the emission line for localized electrons
near the Fermi energy of the electron liquid. { Conversely, no
  such matching of the chemical potentials is expected in
  non-monovalent metals, because there is no continuous process that
  converts a Wigner solid into a lattice with more than one electron
  per site. The resulting mismatch in the chemical potentials actually
  corresponds to the recoil energy of the electrons sharing the site
  with the photoejected electron. Thus we predict that in
  non-monovalent metals, there will {\em also} be a photocurrent due to
  localized sources, but the energy of the outgoing electron will be
  down shifted, relative to the Fermi level, by the said recoil
  energy.}

\begin{figure}[t]
  \centering
  \includegraphics[width=0.75 \columnwidth]{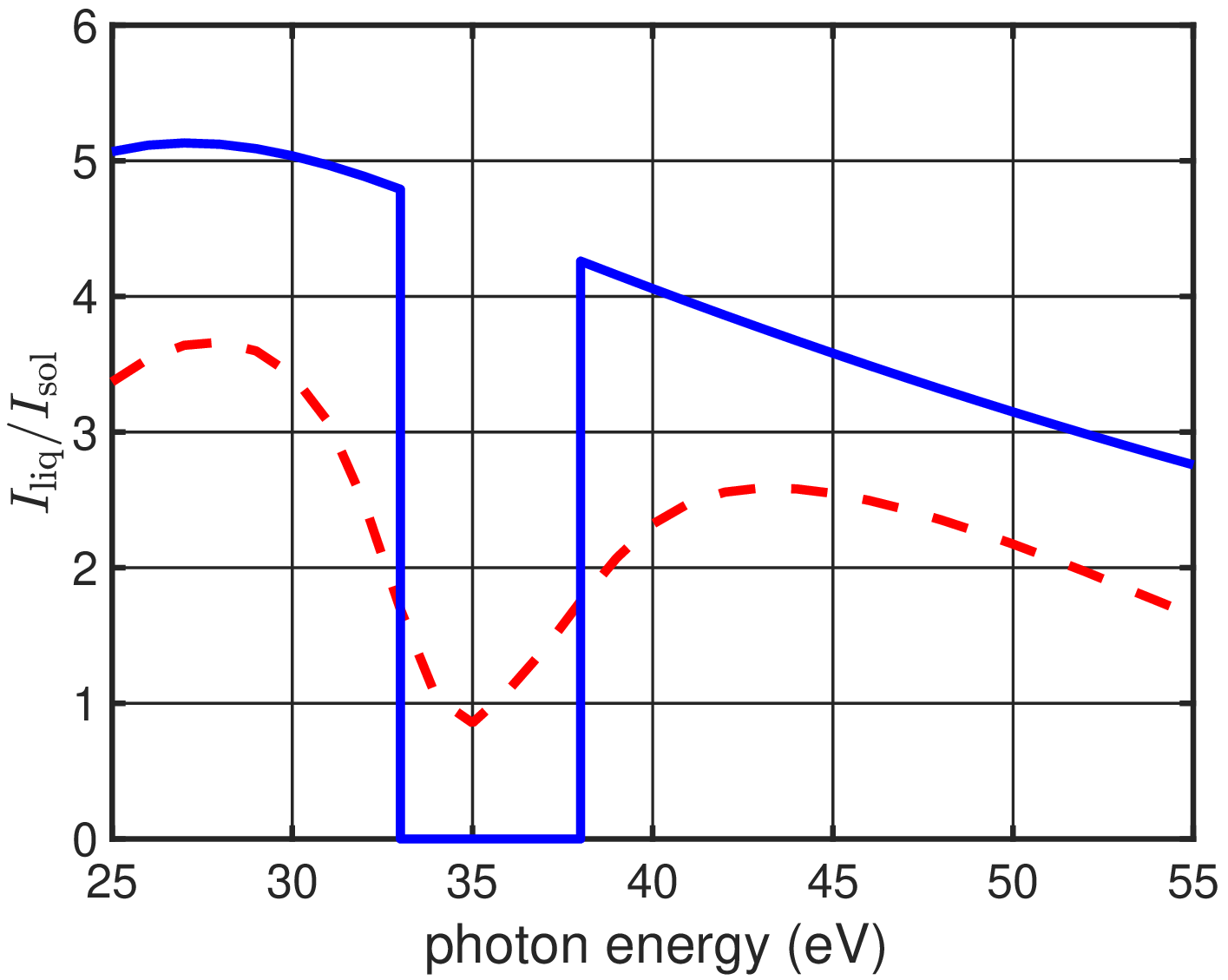}
  \caption{The ratio of the integrated photocurrents from
    Eq.~(\ref{liqsolratio}) as a function of the photon frequency
    without (solid line) and with (dashed line) line broadening
    effects included.}
  \label{IIfig}
\end{figure}

The dispersing part of the photocurrent is, likewise, largely
determined by the Fourier transform of the wavefunction of the atomic
valence shell, but within a near vicinity of the momentum of the
outgoing electron, according to a standard
calculation~\cite{DmitrievThesis, AshNaPseudopot, MahanShungOriginal,
  LLquantum, HufnerPE_book, cohen1991quantum, ABW, Goldenfeld}
detailed in Supplemental Material.~\cite{SMCavitation} When the
broadening of the dispersing peak is neglected, one obtains a rather
simple expression for the relative intensity of the liquid- and
solid-like contributions to the photocurrent, at a given value of the
photon frequency:
\begin{equation} \label{liqsolratio}
  \frac{I_\text{liq}}{I_\text{sol}} = \frac{\pi^{1/2}}{2} \:
  \frac{k^3_{f, \text{liq}}}{k^3_{f, \text{sol}}} \frac{\left|\widetilde
        {\psi}_\text{fr} (k_{f, \text{liq}})\right|^2}{ \int_0^\infty
        \frac{dq}{q} e^{-k_{f, \text{sol}}^2/q^2} \, \left|\widetilde
             {\psi}_\text{fr} (q)\right|^2}.
\end{equation}
Here $k_{f, \text{liq}}$ and $k_{f, \text{sol}}$ denote the momentum
for the outgoing electron extracted as a wave-packet and localized
object, respectively. The two momenta are rather close numerically
because the photon frequency is much greater than $E_F$. According to
Eq.~(\ref{liqsolratio}), the contribution of the localized electrons
is distributed over a broad momentum range and, thus, should be
suppressed but only several-fold relative to the momentum-conserving
transitions, except when the latter transitions fall into the
spectrally forbidden region. The result of the calculation, shown in
Fig.~\ref{IIfig} with the solid line, is consistent with this
expectation.

The peak due to localized sources of photocurrent is intrinsically
broadened owing to the short-lived nature of the effective confining
potential due to the transient electronic solid, as already discussed.
Smaller in magnitude, but significant methodologically is the
broadening of the Fermi-energy peak due to recoil currents of the
electron fluid. These currents must arise because a spatially uniform
fluid is not the ground state of the electron assembly in the presence
of a bounding potential due to the (photo-induced) localized hole.
The recoil currents are entirely analogous to those arising during
photoemission from a deep localized state.  Under the latter
circumstances, a sharp absorption line will broaden to become a skewed
peak, the low-energy side of which is an integrable power
divergence:~\cite{DoniachSunjic, Hopfield}
\begin{equation}  \label{sing}
  I_\text{sol}(E) \propto \frac{1}{\left(E-E_f\right)^{1-\alpha}}
\end{equation}
where
\begin{equation} \alpha = 2 \sum_l (2l + 1) (\delta_l/\pi)^2
\end{equation}
and $\delta_l$ is the phase shift for scattering, due to the
aforementioned local potential, at value $l$ of the angular
momentum. As alluded to already, the majority of scattering in alkali
metals occurs at $l=0$.  If we assume, for simplicity, that the
scattering is exclusively in the $l=0$ channel and that the Friedel
sum rule~\cite{FriedelSumRule} $1 = (2/\pi) \sum_l (2l + 1) \delta_l$
applies, we obtain $\delta_0 = \pi/2$, thus yielding $\alpha = 1/2$.
This is the same exponent for the loss spectrum as in the classical
limit of the Newtonian liquid considered earlier. Consistent with this
notion, the phase shift $\pi/2$ corresponds to a purely viscous
response, whereby for an oscillating signal $e^{i \omega t}$ the
momentum transfer rate goes as $\eta (d/dt) e^{i \omega t} = \eta \,
\omega \, e^{i (\omega t + \pi/2)}$. Still, one should generally
expect scattering at $l > 0$ as well, which will amount to deviations
from the hydrodynamic result in Eq.~(\ref{hydro}).

Although the localized electron is extracted near the absorption edge,
the present situation is distinct from the X-ray edge problem. There,
the excited electron (hole) scatters from a core orbital right into
the continuum perturbed by the excess local potential created by the
excitation; thus the electron (hole) itself contributes to the recoil
currents and, in turn, the overall response of the
fluid.~\cite{NozieresDeDominicis, Hopfield} Here, instead, the
outgoing electron---which had been localized in the first place---does
not itself contribute to the recoil currents. Thus in contrast with
the conventional edge problem, the recoil always results in a
divergence at the spectrum's edge.

We estimate the weights $x_\text{liq}$ and $x_\text{sol}$ in
Eq.~(\ref{Itotal}) by comparing the times valence electrons spend in
classically forbidden and allowed regions, respectively
\begin{equation} \frac{x_\text{liq}}{x_\text{sol}}
  \approx 4 D \frac{t_\text{forbidden}}{t_\text{allowed}}.
 \label{eqTimes1} 
\end{equation}
where we have also included the transmission coefficient $D$ for the
tunneling so as to account only for successful tunneling attempts.
Because the occupied portion of the valence band in alkali metals is
comparable in width to that of a free electron gas at the same
density, one may assume that the potential energy barrier separating
two nearest neighbor sites is sufficiently narrow so that the shape of
the potential energy maximum separating two nearest-neighbor ionic
cores can be well approximated by an inverted parabola. Thus the time
the electron travels one way under the barrier is given by
$t_\text{forbidden} = \pi/\omega^\ddagger$, where $\omega^\ddagger$ is
the frequency of the motion within the parabola. We estimate the
frequency $\omega^\ddagger$ by using the curvature of the Thomas-Fermi
potential at distances corresponding to the midpoint between two
nearest atoms in the lattice. The residence time in the classical
region is determined by the plasmon frequency itself,
$t_\text{allowed} = \pi/\omega_p$, since this is the pertinent
frequency for charge oscillations even on small lengthscales $\sim a$,
in view of the dispersion $\omega_p(k)$ being relatively weak. We
qualitatively estimate the transmission coefficient as the ratio of
the band width for the electrons in the metal and free electrons: $D =
m/m^*$. (We adopt $m^*/m = 1.28$ for the effective mass of the
electron relative to its free-particle value.\cite{PhysRev.118.958,
  PhysRev.132.1991}.) Indeed, the band width scales with the tunneling
matrix element, while one should recover $D = 1$ for free
electrons. The factor 4 reflects that there are 4 inter-nuclear spaces
separating closest neighbors in the BCC lattice per nucleus. This
yields $x_\text{liq}/x_\text{sol} \approx 1.85$.

\begin{figure}[t]
  \centering
  \includegraphics[width=0.9 \columnwidth]{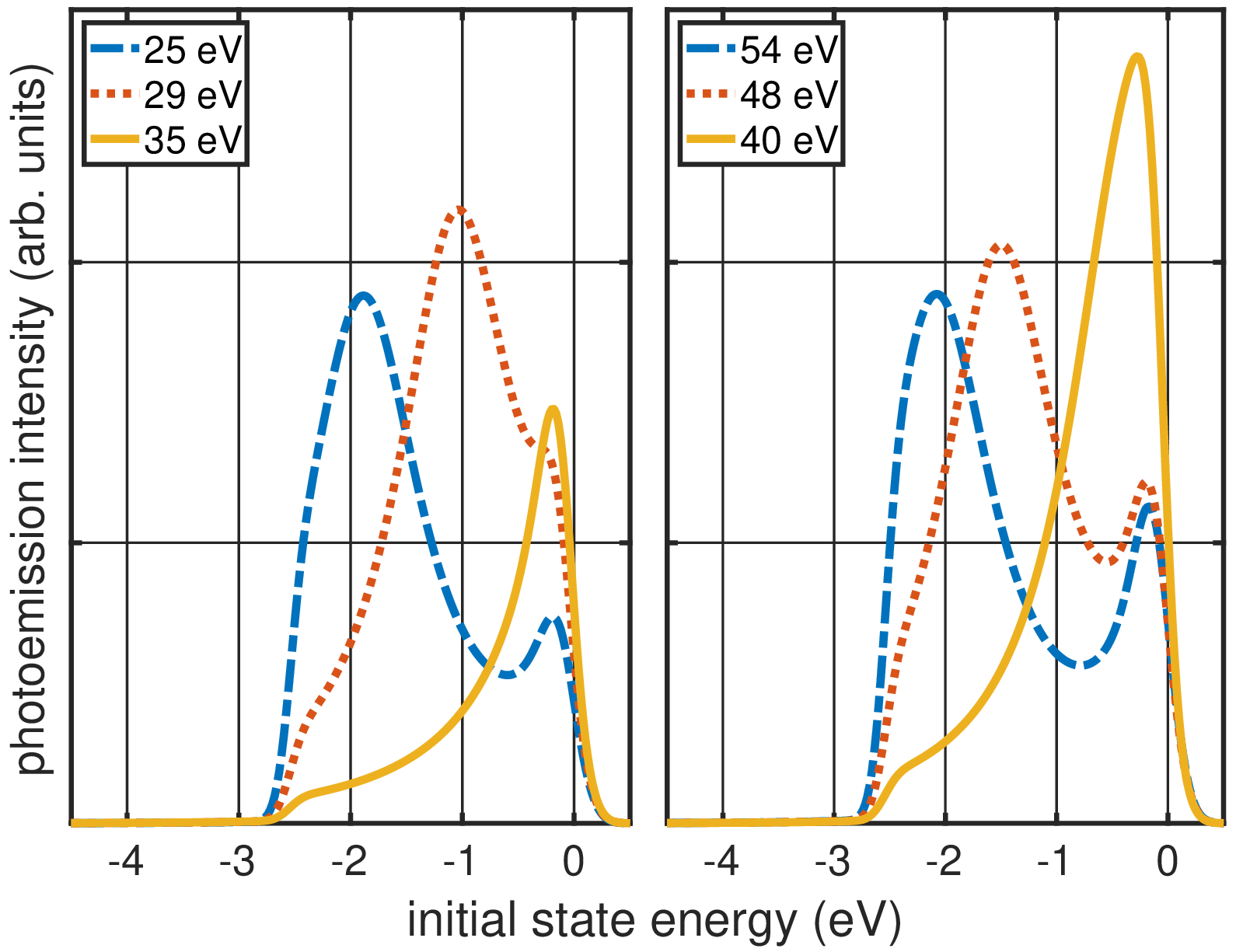}
  \caption{Photoemission spectra from Eq.~(\ref{Itotal}) data for
    sodium in the $(1, 1, 0)$ direction for select values of photon
    frequencies indicated in the legend, to be compared with Fig.~2
    from Ref.~\cite{PlumFirst}. The electron energy is relative to the
    Fermi level. We set $\alpha=1/2$ in Eq.~(\ref{sing}). }
  \label{spectra1}
\end{figure}

{ Thus we have argued that a substantial contribution to the overall
  electronic wavefunction is due to localized electrons. The
  photocurrent due to this contribution, per absorbed photon, is
  comparable to that stemming from the Fermi-liquid. Spectrally, the
  photocurrent due to localized sources is tied to the Fermi level,
  apart from some broadening.  Put together, the above notions then
  rationalize the puzzling Fermi-energy photoemission peak in alkali
  metals.} We evaluate the photoemission spectra for sodium, shown in
Fig.~\ref{spectra1}, to be compared with Fig.~2 of
Ref.~\cite{PlumFirst}. (Experimental spectra also contain a background
due to a variety of processes,~\cite{HufnerPE_book} not considered
here.)  The extent of broadening of the dispersive line was chosen by
hand to be similar to that seen in the experiment; the breadth is
nonetheless consistent with the electron's mean free
path.~\cite{MeanFreePathCalc, BindingEnergyShifts1, HLtime} The
integrated intensity of the Fermi peak is less than that for the
dispersive peak, when the latter is allowed, but the Fermi peak is
also sharper near the top and remains visible within a substantial
range of photon energies.  This is qualitatively consistent with
experiment. { Incidentally we note that the narrowness of the
  anomalous peak, relative to the dispersive peak, is consistent with
  the former stemming from non-itinerant electrons.}  Because the
dispersive peak is broadened, its total intensity is somewhat
diminished in the spectrally allowed range. Conversely, even when the
center of the dispersive peak is in the spectrally forbidden region,
its wings generally extend in the occupied region of the valence band,
as in Refs.~\cite{PhysRevLett.57.1076, PhysRevB.36.4499}. The
intensity ratio from Eq.~(\ref{liqsolratio}) corrected for these
broadening effects is shown in Fig.~\ref{IIfig} with the dashed
line. For the anomalous peak, we adopt the parametrization from
Ref.~\cite{MahanShifts}.

To avoid ambiguity, we note the present analysis pertains
exclusively to the bulk physics. Effects of the free surface are not
included. Also, strictly speaking, on the short time scales of
photoemission, plasmons should be regarded as a symmetry lowering
perturbation to the Fermi-liquid. These effects can be visualized and
contribute to the background,~\cite{DmitrievThesis, CP2KMainReference}
see Supplemental Material.~\cite{SMCavitation}

{ Photoemission out of a localized state exhibits classical features,
  consonant with the infrared
  catastrophe~\cite{AndersonInfraredCatastrophe} accompanying the
  recoil currents. (Similar catastrophes, leading to a classical-like
  localization of a quantum motion, take place as part of the
  Kondo-effect or, for instance, the localization of a particle
  interacting with a bath.~\cite{SB_review}) The photon's momentum and
  energy are initially imparted to the material within a localized
  region, before they are passed on to the rest of the electrons and
  the nuclei. The density of the grand-canonical free energy
  ($V=\text{const}$, $T=\text{const}$) is equal to the negative
  pressure.~\cite{CL_LG} Thus a photon impinging on a volume $V$
  effectively creates an excess negative pressure $-\hbar
  \omega_\text{ph}/V$, the corresponding forces being much greater
  than the characteristic electronic forces when $V \lesssim a^3
  (\hbar \omega/E_F) \simeq 10^1 a^3$.  This negative pressure is, in
  fact, the driving force behind the decrease in local density caused
  by photoemission. This negative pressure will persist until the hole
  is filled by the recoil currents. When creating a localized hole, we
  are brining in physical contact two distinct phases thus incurring a
  mismatch penalty. (If the interface is thin, the penalty amounts to
  a conventional surface tension.~\cite{CL_LG}) Thus the formation of
  the hole is a {\em nucleation}-like process. Nucleation of a
  low-density phase caused by local excess of negative pressure is a
  well known phenomenon called ``cavitation,'' hence the title of the
  article.  Nucleation, of course, is a strongly non-linear
  phenomenon; {\em homogeneous} nucleation furthermore represents a
  breaking of spatial symmetry.}

As a dividend of the present analysis, we note that the metal acts as
a transmitter of electrons, during photoemission, the final state of
the particle being a plane wave.  Conversely, one may consider the
reciprocal process, in which the metal is the receiver. Hereby the
electron is initially a plane wave and eventually enters the metal in
the form of a localized object while filling a localized vacancy,
whose concurrent formation is accompanied by recoil currents.  This
reciprocal process can be thought of as a detection event for the
received electron, so that its location becomes known to the extent
determined by the size and shape of the hole.  As the measurement
proceeds, the single-particle density matrix progressively deviates
from its initial, nearly free-electron value. Thus the present
scenario is an explicit example of how a measurement event is a result
of a strongly non-linear, many-body phenomenon but can be profitably
thought of as a ``collapse'' of a one-particle wavefunction, if the
full density-matrix of the system is unavailable. Because the event is
an instance of symmetry breaking, its outcome is history dependent. {
  This is analogous to how the precise magnetization pattern in a
  magnet below the Curie point or, to give another example, the
  detailed structure of a glass below the glass transition~\cite{L_AP}
  depend on the preparation protocol.} Other, arguably simpler
scenarios can be imagined. For example, consider a dilute gas or a set
of deep core orbitals in a metal such that the photon's wavelength is
much greater than the spatial separation between pertinent
orbitals. An electron to be photoemitted is automatically localized
within an orbital whose eventual identity will have emerged as a
result of a symmetry lowering process and, thus, depends on the
history. The non-linear phenomenon is whatever the process that causes
individual orbitals not to form a band in the first place, such as a
metal-insulator transition.~\cite{LKgap} Likewise, receiving orbitals
are also automatically localized.

\subsection*{Acknowledgments} We thank E. Ward Plummer (now deceased) and
Gerry Mahan (now deceased) for inspiring conversations and Peter
G. Wolynes and Eric Bittner for helpful insight.  We gratefully
acknowledge the support by the NSF Grants CHE-1465125 and CHE-1956389,
the Welch Foundation Grant E-1765, and a grant from the Texas Center
for Superconductivity at the University of Houston. We gratefully
acknowledge the use of the Maxwell/Opuntia Cluster at the University
of Houston.  Partial support for this work was provided by resources
of the uHPC cluster managed by the University of Houston and acquired
through NSF Award Number ACI-1531814.

\bibliography{/Users/vas/Documents/tex/ACP/lowT,pes}

\clearpage

\setcounter{section}{0}
\setcounter{equation}{0}
\setcounter{figure}{0}
\setcounter{table}{0}
\setcounter{page}{1}
\makeatletter
\renewcommand{\theequation}{S\arabic{equation}}
\renewcommand{\thefigure}{S\arabic{figure}}
\renewcommand{\thetable}{S\arabic{table}}
\renewcommand{\bibnumfmt}[1]{[S#1]}


  \begin{center} 
  
  { \bf \large {\em Supplemental Material}\/: Cavitation in Electron Fluids and \\ the Puzzles of
  Photoemission Spectra in Alkali Metals} \medskip
  
  {\large Roman Dmitriev$^{1}$, Jenny Green$^{2}$, and Vassiliy
      Lubchenko$^{1, 3, 4}$} \medskip
      
    {\normalsize $^1$Department of Chemistry, University
  of Houston, Houston, TX 77204-5003

 $^2$St. John's School, Houston, TX 77019

     $^3$Department of Physics, University of Houston, Houston, TX
  77204-5005
  
  $^4$Texas Center for Superconductivity,
  University of Houston, Houston, TX 77204-5002}

%
%
  
%
%
%

  \end{center}
  


This Supplemental Material is organized as follows: In Section
\ref{photocurrent}, we detail the present estimates for the
photocurrent due to both wavepacket-like and localized sources.  In
Section~\ref{fermi}, we illustrate effects of transient symmetry
lowering due to plasma oscillations.

\section{Calculation of Photocurrent}
\label{photocurrent}

We quantify the rate of photoemission using an effective one-particle
description, whereby the initial state of the electron is described by
an energy $\varepsilon_i$ and wavefunction $\psi_i$, while in the
final state, the electron's energy is $\varepsilon_f$ and the
wavefunction $\psi_f$.  We evaluate the rate $w_{fi}$ of transitions
from the initial to the final electronic state, in the presence of a
photon at frequency $\omega_\sph$, using the Fermi Golden
rule:~\cite{LLquantum, HufnerPE_book}
\begin{equation}
  \label{eqwfi1} 
  w_{fi} = \frac{2\pi}{\hbar}| M_{f, i} |^2\delta(\varepsilon_f -
  \varepsilon_i - \hbar\omega_\sph),
\end{equation}
where 
\begin{equation}
\label{eqMfi1} 
M_{f, i}=\bra{\psi_f} H^{'} \ket{\psi_i}
\end{equation}
is the one-particle transition matrix element for the perturbation
$H^{'}$ due to the photon field. In the above expression, we have
omitted the overlap of the rest $(N-1)$ wavefunctions for the
electrons that are left behind in the metal, following a photoemission
event. This overlap generally differs from unity. The corresponding
effects---along with those stemming from correlations---modify the
line shape of the photoemission spectra, to be discussed in due time.

The photoemission spectrum $I(E, \hbar\omega_{p})$ is the total
current per unit of electron energy $E$:
\begin{equation}
\label{eqIinteg1} 
I(E, \hbar\omega_\sph) = \frac{2\pi}{\hbar} \sum_{i, f} |M_{f, i}|^2
\, \delta(\varepsilon_f - \varepsilon_i - \hbar\omega_\sph) \,
\delta(E - \varepsilon_f - \Phi),
\end{equation}
where one sums over all possible initial and final states of the
electron and the quantity $\Phi$ is the work function of the metal.

In angle-resolved photo-emission spectroscopy (ARPES), one
determines the total amount of collected electrons per energy $E$ and
solid angle $\Omega$. Hereby, one computes the photocurrent for
electrons exiting the sample within a specified solid angle
$\Omega_d$:
\begin{equation}
\label{eqIarpes1} 
I(E, \Omega, \hbar\omega_\sph) = \frac{2\pi}{\hbar} \sum_{i, f } |M_{f,
  i}|^2 \, \delta(\varepsilon_f - \varepsilon_i - \hbar\omega_\sph) \,
\delta(E - \varepsilon_f - \Phi) \, \delta(\Omega - \Omega_d),
\end{equation}
In the remainder of this Supplemental Material (SM), we will drop the
multiplicative constant $2\pi/\hbar$,  to declutter the formulas.

We limit ourselves to the common case of the outgoing electron's
energy being much greater than the crystal field, so that the latter
can be regarded as a weak perturbation. Thus we assume the electron's
energy is a continuous quantity that obeys the dispersion relation for
a free electron, as in Ref.~\cite{PlumFirst}:
\begin{equation} \label{disperout}
  \varepsilon_f(\bk) = \hbar^2 k^2/2 m.
\end{equation}
Consequenctly, the summation over final states can be rewritten as a
continuous integration over the wave vector of the outgoing electron:
\begin{equation}
\label{eqIarpes3}
I_{\Omega}(E, \Omega, \hbar\omega_\sph) = \sum_{i}\; \int d^3 \bk
|M_{f, i}|^2 \, \delta(\Omega(\bm k) - \Omega_d) \,
\delta(\varepsilon_f(\bm k) - \varepsilon_i - \hbar\omega_\sph) \,
\delta(E - \varepsilon_f(\bm k) - \Phi)
\end{equation}

We will suppose, in the conventional fashion, that the photoemitted
electrons are collected in the direction normal to the surface of the
sample; for instance in Jensen and Plummer's
experiment,~\cite{PlumFirst} this direction corresponds to the normal
of the crystal plane $(1, 1, 0)$. Thus we obtain
\begin{equation}
\label{eqIarpes4}
I_{\Omega}(E, \Omega, \hbar\omega_\sph) = \sum_{i}\; \int d k \, k^2 
|M_{f, i}|^2 \, 
\delta(\varepsilon_f(\bm k) - \varepsilon_i - \hbar\omega_\sph) \,
\delta(E - \varepsilon_f(\bm k) - \Phi)
\end{equation}

Integration with respect to the momentum of the outgoing electron then
yields
\begin{equation}
\label{eqIarpes6}
\begin{split}
I (E, \hbar\omega_\sph) & = \frac{ {k_{f}}^2 } {\left| d
  \varepsilon_f(k)/dk \right|_{k_f} } \sum_{i}\; \left| M_{f, i}
\right|^2_{\bk_f} \delta(E - \varepsilon_f(\bm k_f) - \Phi)\\ & =
\frac{m \, k_{f}}{\hbar^2} \sum_{i}\; \left| M_{f, i}
\right|^2_{\bk_f} \delta(E - \varepsilon_f(\bm k_f) - \Phi)
\end{split}
\end{equation}
where the value of the outgoing momentum is fixed according to
\begin{equation} \label{enconservout}
  \varepsilon_f(\bm k_f) - \varepsilon_i = \hbar\omega_\sph
\end{equation}
so as to satisfy energy conservation, as embodied by the first delta
function in Eq.~(\ref{eqIarpes4}).

To evaluate the transition matrix element we use the expression for
the energy of an electron in the presence of electromagnetic field
with the vector potential $\bm A$:\cite{cohen1991quantum}
\begin{equation}
  \label{eqDip2} 
  H = \frac{(\bm p - q_e \bm A)^2}{2m} + U \equiv H_0 + H',
\end{equation}
where $q_e$ is the charge of the electron and $H'$ is the perturbation
due to the photon field, by construction. In the limit of linear
response, one obtains:
\begin{equation}
  \label{eqDip3} 
  H' = - \frac{q_e}{2m}(\bm p\bm A + \bm A \bm p) = - \frac{q_e}{m} \bm A
  \bm p = i \hbar \frac{q_e}{m} \bm A \nabla,
\end{equation}
where we used $\bm p = - i\hbar\bm\nabla$ and $\bm p\bm A =
-i\hbar\bm\nabla\bm A = -i\hbar\left[(\bm\nabla\cdot\bm A) + \bm
  A\bm\nabla\right] = -i\hbar\bm A\bm\nabla = \bm A\bm p$, since in
the Coulomb gauge $(\bm\nabla\cdot\bm A)=0$. Subsequently,
\begin{equation}
\label{eqDip6} 
M_{f, i} \equiv \bra{\psi_f}H^{'}\ket{\psi_i} = -\frac{q_e}{m}
\bra{\psi_f}{\bm A} \bm p\ket{\psi_i}
\end{equation}
where, specifically, the photon field is a plane wave
\begin{equation} \label{Aphoton} \bm A =
  \bm e A_0 e^{i \bm k_\sph \bm r}.
\end{equation}
Here $\bm k_\sph$ denotes the photon's momentum and $\bm e$ is the
unit vector specifying the direction of the vector potential.
Consequently, Eq.~(\ref{eqDip6}) yields
\begin{equation}
\label{eqLoc4}
|M_{f, i}|^2 = \left(\frac{q_e \hbar}{m}\right)^2|\bra{\psi_f(\bm r)}\bm
A \bm\nabla \ket{\psi_i(\bm r)}|^2 = \left(\frac{q_e
  \hbar}{m}\right)^2|\bra{\psi_i(\bm r)}\bm A^* \bm\nabla
\ket{\psi_f(\bm r)}|^2.
\end{equation}

The final-state wavefunction is a plane wave, as already alluded to:
\begin{equation}
\label{eqLoc3}
\psi_f(\bm r) = e^{i\bm k_f \bm r}
\end{equation}
---where we do not need concern ourselves with the normalization---and
so
\begin{equation}  \label{dps}
  \bm\nabla \psi_f(\bm r)  = i \bk_f \psi_f(\bm r).
\end{equation}

In what follows, we consider two cases: when the initial state is a
Bloch wave and a localized entity, respectively.

\subsection{Photocurrent due to Bloch states}

We use the standard expression for a normalized Bloch
wave:\cite{Ashcroft}
\begin{equation}
\label{eqLoc1}
\begin{split}
& \psi_i(\bm r)  = \frac{1}{\sqrt{V}} \phi_{\bk_i}(\bm r)e^{i \bk_i
  \br} \\ & \int\displaylimits_{V} d^3 \br |\psi_i (\br)|^2 = 1
\end{split}
\end{equation}
where the function $\phi_{\bk_i}(\br)$ has the periodicity of the crystal
lattice and does not depend on the sample's size.  Consequently,
Eqs.~(\ref{eqLoc4}) and (\ref{dps}) yield
\begin{equation}
  \label{eqLoc5}
|M_{f, i}|^2 = \frac{1}{V} \left(\frac{q_e A_0\hbar}{m}\right)^2 \left|
\int\displaylimits_{V} d^3 \bm r \, e^{-i(\bk_f - \bk_i - \bk_\sph)\bm
  r} \phi_{\bk_i}(\bm r) \right|^2 (\bk_f \bm e)^2
\end{equation}
where we used Eqs.~(\ref{Aphoton}) and (\ref{eqLoc3}).

To make explicit the role of periodicity of the function
$\phi_{\bk_i}(\bm r)$---as stemming from the periodicity of the
underlying crystal, of course---we present the Bloch state
$\phi_{\bk_i}(\bm r)$ in Eq.~(\ref{eqLoc5}) as the following sum:
\begin{equation} \label{phiper}
  \phi_{\bk_i}(\bm r) = \sum_j^N \phloc(\br-\br_j, \bk_i)
\end{equation}
where the points $\br_j$ form a periodic array that mirrors the
periodicity of the underlying crystal. It is convenient to think of
space as broken up into cells, so that there is exactly one point
$\br_j$ per cell. The choice of the function $\phloc$ is not unique,
of course, and is made according to one's convenience. Any two
functions $\phloc(\br-\br_j, \bk_i)$ pertaining to two distinct
centers $\br_j$ can overlap; for instance, a plane wave corresponds to
$\phi(\br-\br_j, \bk_i) = \text{const} = 1/N$. Still, we ordinarily
have in mind such functions $\phloc(\br, \bk_i)$ whose magnitude
rapidly decays away from the origin at sufficiently large distances,
as would atomic wavefunctions.  In any event, one may now present the
integral over space in Eq.~(\ref{eqLoc5}), as a sum of integrals of
individual functions $\phloc(\br, \bk_i)$:
\begin{equation}
\label{eqLoc6}
|M_{f, i}|^2 = \frac{1}{V} \left(\frac{q_e A_0\hbar}{m}\right)^2
\left| \sum_j e^{-i(\bk_f - \bk_i - \bk_\sph) \br_j} \int d^3 \br \,
e^{-i(\bk_f - \bk_i - \bk_\sph) (\br-\br_j)} \phloc(\br-\br_j, \bk_i)
\right|^2 (\bk_f \bm e)^2.
\end{equation}

For a large system, the summation over the cells in Eq.~(\ref{eqLoc6})
singles out only such values of the vector $(\bk_f - \bk_i -
\bk_\sph)$ that are each equal to vectors of the reciprocal lattice,
in which case the sum is equal to the number itself of the cells,
i.e., $V/v_c$, where $v_c$ is the cell's volume. For all other values
of that vector, the sum scales sub-thermodynamically, which then
amounts to momentum conservation, up to Umklapp processes. One thus
obtains:
\begin{eqnarray}
\label{eqLoc7.1}
|M_{f, i}|^2 & = & \frac{V}{v_c^2} \left(\frac{q_e
  A_0\hbar}{m}\right)^2 (\bk_f \bm e)^2 \sum_{\bm G} \left| \ptloc(\bm
G, \bk_i) \right|^2 \delta_{\bk_f - \bk_i - \bk_\sph, \bm G}
\\ \label{eqLoc7.2} &\approx & \frac{V}{v_c^2} \left(\frac{q_e
  A_0\hbar}{m}\right)^2 (\bk_f \bm e)^2 \sum_{\bm G} \left| \ptloc(\bm
G, \bk_i) \right|^2 \delta_{\bk_f - \bk_i, \bm G}
\end{eqnarray}
where we define the Fourier transform of the function $\phloc$ as
follows:
\begin{equation} \label{locFourier} \ptloc(\bq, \bk_i)
  = \int d^3\br \: \phloc(\br, \bk_i) \: e^{-i\bq \br}
\end{equation}
The discrete sum in Eqs.~(\ref{eqLoc7.1}) and (\ref{eqLoc7.2}) runs
over all vectors $\bm G$ of the lattice reciprocal to the lattice
comprised by the cells.  The approximate equality in
Eq.~(\ref{eqLoc7.2}) holds as long as the photon's wave length is much
greater than the unit cell---whereby $k_\sph \ll G$---which is the
case at photon's energies in Plummer et al.'s
studies.~\cite{1402-4896-1987-T17-021} The delta-function in
Eqs.~(\ref{eqLoc7.1}) and (\ref{eqLoc7.2}) is of the Kronecker
variety, so that summation over any of its arguments yields exactly
{\em unity}, $\delta_{\bk, \bk} =1$, while, at the same time, singling
out the appropriate value of the argument in the rest of the
expression. The set of the reciprocal vectors $\bm G$ contains each
vector exactly {\em once} and the summation over $\bm G$ is already
explicitly present in the expression. Consequently, if one must sum
over any of the electronic momenta, this can be done via continuous
integration using weight functions normalized to unity, not
volume. Thus we explicitly see the amount of photocurrent scales
linearly with the sample's volume, as it should. One may equivalently
employ the Dirac delta-function instead of the Kronecker variety,
using the conventional procedure explained, for instance, in Section
5.7.2 of Ref.~\cite{Goldenfeld}: $V \delta_{\bk_1, \bk_2} \to (2
\pi)^3 \delta(\bk_1 - \bk_2)$ while summation, if any, over electronic
momenta must be replaced by continuous integration according to the
prescription $\sum_k \to V \int d^3 \bk/(2 \pi)^3$. The latter
procedure however makes the thermodynamic scaling of the photocurrent
less explicit.

Expression (\ref{eqLoc7.2}) indicates that an electron to be
photoejected must be a Bloch state, not a plane
wave---$\phi_{\bk_i}(\br) \ne \text{const}$---in order for the
integral over the cell to be non-vanishing; this is consistent with
the general notion that a free electron can not absorb or emit a
photon. To choose the appropriate approximation for the wavefunction
that enters the rather general expression (\ref{eqLoc7.2}), it is
useful to survey the range of possible responses of the electron fluid
in terms of the magnitude of the recoil momentum. We begin from the
low-momentum limit, in which one formally assumes the molecular
potential is only weakly non-uniform; this appears to be the most
common approximation.~\cite{AshNaPseudopot, MahanShungOriginal} Hereby
one focuses on the lowest order harmonics of the spatial variation the
crystal field, which correspond to the spacing between nearest
neighbors in the lattice.  Using the following commutation
relationship for the unperturbed single electron Hamiltonian $H_0$:
\begin{equation}
\label{eqMfiDip1} 
\left[H_0, \bm p\right] = i\hbar\bm\nabla U
\end{equation}
we obtain:
\begin{equation}  i\hbar \bra{\psi_f} \bm\nabla U \ket{\psi_i} =
  \bra{\psi_f} \left[H_0, \bm p\right] \ket{\psi_i} = (E_f - E_i)
  \bra{\psi_f} \bm p \ket{\psi_i},
\end{equation}
where $E_i$ and $E_f$ are the pertinent eigenvalues of the unperturbed
Hamiltonian $H_0$.  Eqs.~(\ref{eqDip6}) and (\ref{Aphoton})
consequently yield that the transition matrix element is directly
related to the corresponding matrix element of the force due to the
crystal field:
\begin{equation}
M_{f, i} = - \frac{i q_e A_0\hbar}{m(E_f - E_i)} \bra{\psi_f}\bm e
 \bm \nabla U\ket{\psi_i} = - \frac{iq_e A_0}{m\omega_\sph} \bra{\psi_f}\bm
e \bm \nabla U\ket{\psi_i}
\end{equation}
and in the second equality we substituted $E_f - E_i = \hbar
\omega_\sph$, in view of Eq.~(\ref{eqIinteg1}). Regarding the
molecular potential as a perturbation, the lowest order approximation,
in terms of $\nabla U$, is obtained by setting $\phi_{i} (\br) = 1$.
Repeating the steps leading to Eq.~(\ref{eqLoc7.2}), one thus obtains
for the square of the transition moment, $|M_{f, i}|^2$, from
Eq.~(\ref{eqIarpes1}):
\begin{equation}
\label{Ubased}
|M_{f, i}|^2 \approx V \left(\frac{q_e A_0 \hbar}{m}\right)^2 \sum_{\bm G} (
\bm e \bm G)^2 \frac{|\widetilde U_{\bm
    G}|^2}{(\hbar \omega_\sph)^2} \delta_{\bk_f - \bk_i, \bm G}
\end{equation}
where the quantity
\begin{equation}
\label{Uk}
\widetilde  U_{\bk} \equiv \int\displaylimits_{\text{cell}}
\frac{d^3\bm r}{v_c} \, e^{-i \bk \br} \: U (\bm r)
\end{equation}
is the (discrete) Fourier component of the molecular field. Note the
integration in Eq.~(\ref{Uk}) is over the cell, not over the whole
space, c.f. Eq.~(\ref{locFourier}). 

In view of the long wavelength nature of the approximation leading to
expression (\ref{Ubased}), it seems difficult to ascertain the
accuracy of the large-$G$ terms in the sum over $\bm G$; thus it seems
prudent to include only the lowest-$G$ term in the sum.

In constrast, the large $\bk_f$ asymptotics of the photocurrent are
largely determined by the behavior of individual atomic wavefunctions
near their respective nuclei. This is where the rate of spatial
variation of the wave function reaches its maximum value, owing to the
intense scattering of the electron off the largely unscreened
potential due to the nucleus. To see this explicitly, we assume that
in an alkali metal, only one frontier atomic orbital, call it
$\psi_\text{fr} (\br)$, contributes to the Bloch states. (This
orbital, approximately, would be the 3s atomic orbital for Na or 4s
orbital for K.)  Hence, the following ansatz approximates the Bloch
state of the initial state near the nuclei, the approximation being
worse in the inter-nuclear space:~\cite{ABW}
\begin{equation} 
  \psi_i(\bm r) = \frac{1}{\sqrt{N}} \sum_{\bR} e^{i \bk \bR } \, \psi_\text{fr}
  (\br-\bR)
\end{equation}
where the summation is over the locations of the atomic cores. In view
of Eqs.~(\ref{eqLoc1}) and (\ref{phiper}), the ansatz above implies
$\phloc(\br, \bk_i)=v_c^{1/2} e^{-i \bk_i \br}
\psi_\text{fr}(\br)$. Combined with momentum conservation, as
expressed by the delta function in Eq.~(\ref{eqLoc7.2}), this yields:
\begin{equation} \label{FT}  \ptloc(\bm
G, \bk_i) \approx v_c^{1/2} \: \widetilde{\psi}_\text{fr} (\bk_f)
\end{equation}
where we specifically adopt the primitive cell as the cell in
Eq.~(\ref{eqLoc7.1}), so that $V/N = v_c$, and define the Fourier
transform of the frontier orbital in the usual fashion:
\begin{equation} \label{FTphifr}
\widetilde{\psi}_\text{fr} (\bk) = \int d^3\br \, e^{-i \bk \br}
\psi_\text{fr}(\bm r).
\end{equation}

Thus the transition matrix element is approximately given by the
Fourier transform of the frontier orbital at the momentum $\bk_f$ of
the outgoing electron.  The latter wavefunction has a singularity at
the origin in the form of a discontinuity in the first derivative, due
to scattering off the nucleus. Thus the leading large $q$ asymptotics
of the Fourier transform $\widetilde{\psi}_\text{fr} (q)$ have the
form of a power law decay. The magnitude of the tail is largely
determined by the magnitude of the electron's wave function near the
nucleus.  Indeed, the atomic function in question corresponds to a
vanishing orbital momentum, $l=0$. Thus the radial portion of the
wavefunction $\psi_\text{fr} (\br)$ is given by a product of a
relatively slow function, finite at the origin, that determines the
nodal structure of the wavefunction and an exponentially decaying
function that details the tunneling tail at large separations from the
nucleus.~\cite{LLquantum} A straightforward calculation shows that the
aforementioned power law tail, up to a factor of order one, is given
by the expression
\begin{equation} \label{FThighq}
  \widetilde {\psi}_\text{fr} (q) \xrightarrow[q \to \infty]{} 4 \pi
  \, \frac{ \psi_\text{fr} (0)}{L \, q^4}.
\end{equation}
where the quantity $L$ reflects the gross rate of decay of the
wavefunction away from the center. The expression above corresponds to
the following, simplified form for the wavefunction near the nucleus:
$\psi_\text{fr} (\br) \to \psi_\text{fr} (0) \, e^{-r/L}$.

\begin{figure}[t]
  \centering
  \includegraphics[width=0.7\textwidth]{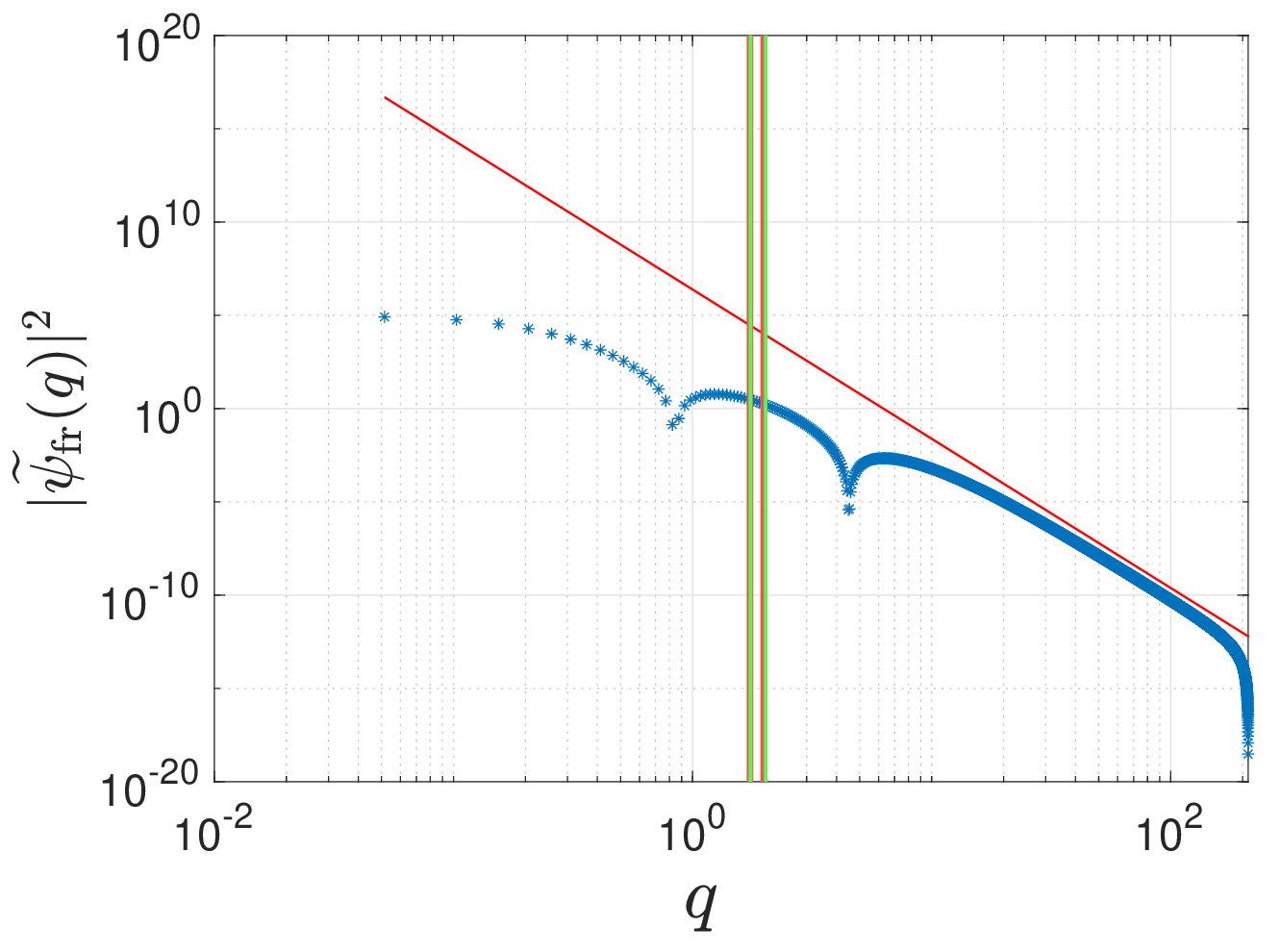}
  \caption{Symbols: Square-modulus of the Fourier transform of the
    (numerically obtained) 3s Thomas-Fermi orbital of Na, as a
    function of the wavevector. The latter is given in atomic
    units. The thin solid red line indicates the slope of the power
    law dependence $1/q^8$. The two sets of vertical lines indicate
    the ranges of the momentum of the outgoing electron for the
    dispersive and non-dispersive peaks pertaining to Fig.~2 of the
    main text. The rapid decay on the r.h.s. of the $q$-range is due
    to the numerical error caused by the finite grid size.}
  \label{FTfig}
\end{figure}

In experiments of Plummer and coworkers, $k_f \sim 3 k_F$. That is,
the integral in Eq.~(\ref{FT}) probes spatial variations in the atomic
function on length scales considerably less than the inter-atomic
spacing but not quite in a regime where the large-$q$ asymptotics from
Eq.~(\ref{FThighq}) set in.  We illustrate this by displaying the
Fourier transform of the 3s Thomas-Fermi orbital for a standalone
sodium atom in Fig.~\ref{FTfig}, where we indicate the ranges for the
pertinent values of the momentum of the outgoing electron by sets of
vertical lines. These pertinent values of momenta are separated by
nodes from both the long-range and short-range behaviors of the
wavefunction that are characteristic of the electronic motions,
respectively, in the inter-nuclear space and near an unscreened
nucleus. In the spirit of tight-binding approximations, we conclude
that using the atomic function for $\psi_\text{fr}$ may lead to a
quantitative, but not qualitative error. Furthermore, this
approximation will allow us to express the most important prediction
of the present treatment, Eq.~(\ref{liqsolratio}), in terms of a ratio
of quantities pertaining to the same object, which should mitigate the
error, if any. In any event, we obtain for the matrix element, in view
of Eq.~(\ref{eqLoc7.1})
\begin{equation}
\label{Mextended}
\begin{split}
|M_{f, i}|^2 & \approx \frac{V}{v_c} \left(\frac{q_e
  A_0\hbar}{m}\right)^2 (\bk_f \bm e)^2 \left| \: \int d^3\br \, e^{-i
  \bk_f \bm r} \psi_\text{fr}(\bm r) \right|^2 \sum_{\bm G}
\delta_{\bk_f - \bk_i, \bm G} \\ &= \frac{V}{v_c} \left(\frac{q_e
  A_0\hbar}{m}\right)^2 (\bk_f \bm e)^2 \left| \widetilde
\psi_\text{fr}(\bk_f) \right|^2 \sum_{\bm G} \delta_{\bk_f - \bk_i,
  \bm G}
\end{split}
\end{equation}

In the assumption of the initial electronic state being a
single-particle state, subject to a stationary crystal field, the
contribution of each vector $\bm G$ to the sums in Eqs.~(\ref{Ubased})
and (\ref{Mextended}) is a sharp peak centered at energy $(\Phi +
\varepsilon_i + \hbar\omega_\sph)$, per Eq.~(\ref{eqIarpes6}). This
notion, combined with the assumption on the dispersion of the outgoing
electron, Eqs.~(\ref{disperout}) and (\ref{enconservout}), as well as
momentum conservation, allows one to determine the one-particle
spectrum $\varepsilon_i(\bk_i)$ in the solid. In the presence of
interactions, however, single-particle wavefunctions are no longer
solutions of the Hamiltonian. As a result, the lifetime $\tau$ of a
quasi-particle is intrinsically finite, which leads to a broadening of
the photoemission line, in addition to the broadening due to 
factors than include, among others, lattice imperfections and finite
spectral resolution of the spectrometer. In turn, this lifetime can be
expressed through the mean free path $l_\text{mfp}$ and the group
velocity $\bm v(\bk) = \partial \varepsilon/\partial (\hbar \bk)$ of
the wavepacket:
\begin{equation} \label{de0}
  \delta \epsilon_i \approx \frac{\hbar}{\tau} \approx |\bm \nabla_k
  \varepsilon_i(\bk_i)| \frac{1}{l_\text{mfp}},
\end{equation}
within a factor of order one.~\cite{HLtime} The approximation
underlying the simple result above can be discussed in a more formal
vein: In the presence of many-body effects, single particle
excitations are described using the (time-dependent) one-particle
density matrix $\rho(\br_1, \br_2)$.  One can Wigner transform this
object: $\rho(\br, \bk) = \int d^3 (\br_2 - \br_1) \rho(\br_1, \br_2)
e^{-i \bk (\br_2 - \br_1)}$, take the limit of slow variation in terms
of the center-of-mass variable $\br=(\br_1+\br_2)/2$, and then Fourier
transform in time. The resulting object, call it $\tilde \rho(\bk,
\omega)$, can be thought of as a distribution of the effective energy
and momentum of the initial state of the quasiparticle. This
distribution is maximized along a surface $\omega_0=\omega_0(\bk_0)$
in the four dimensional space formed by the three components of the
wave-vector $\bk$ and the frequency $\omega = \varepsilon/\hbar$. The
latter surface plays the role of the dispersion relation while the
derivative $\partial \omega_0/\partial \bk_0$ corresponds with the
group velocity of the quasiparticle, in the usual fashion.  Momentum
conservation in Eqs.~(\ref{eqLoc7.1}) and (\ref{eqLoc7.2}) represents
a rigid geometrical constraint, due to the periodicity, on which
spatial harmonics of the (many-body) electronic wavefunction can
contribute to the photocurrent, see also
Ref.~\cite{PhysRevLett.58.959} Deviations from strict periodicity, if
mild, can be effectively incorporated into the treatment by replacing
the delta functions, in the momentum space, by sharply peaked
functions of finite, small width. In any event, the energy of the
initial state remains effectively distributed even if momentum is
strictly conserved, the distribution given by $p(\omega) \propto
\tilde \rho(\bk, \omega)|_{\bk=\bk_f}$. The width of the energy
distribution can be evaluated by multiplying the uncertainty in the
momentum of the wavepacket $\delta p \equiv \hbar \delta k$ by the
rate of dispersion in the direction of steepest descent $\bm \nabla_k
\varepsilon = \bm \nabla_p (\bm p^2/2 m^*) = \bm p/m^*$, the latter
quantity being the group velocity itself, of course.  The inverse of
the wavevector uncertainty is simply the spatial extent $1/\delta k$
of the wavepacket, i.e., the mean free path; thus $\delta k \approx
1/l_\text{mfp}$. The corresponding uncertainty in the energy can be,
then, approximately evaluated as follows:
\begin{equation} \label{de}
  \delta \epsilon_i \approx |\bm \nabla_k \varepsilon_i(\bk_i)| \,
  \delta k \approx |\bm \nabla_k \varepsilon_i(\bk_i)|
  \frac{1}{l_\text{mfp}},
\end{equation}
as before. Already the simple parabolic dispersion relation works well
in Na and K, in which the Fermi surface is spherical---to a very good
approximation---and does not touch the boundaries of the Brillouin
zone:
\begin{equation} \label{disp}
  \varepsilon(\bk_i) = \hbar^2 k_i^2/2 m^*
\end{equation}
where we dropped the subscript ``0'' for typographical clarity and
$m^*$ is the effective mass of the electron. Here we use $m^*/m =
1.28$, as in Ref.~\cite{PlumFirst}. This number happens to be close
numerically to the midway value between the effective mass determined
using calorimetry~\cite{PhysRev.118.958} and cyclotron
resonance,~\cite{PhysRev.132.1991} respectively.

Thus we conclude that for a given value of the momentum $\bk_i$ of the
quasiparticle, the corresponding effective energy is distributed
according to some (normalized) weight function
$w_\text{liq}(\varepsilon_i - \varepsilon(\bk_i), \delta
\varepsilon_i) $:
\begin{equation} \label{wliq} \int dx \: w_\text{liq}(x, \delta \varepsilon_i) = 1,
\end{equation}
In practical terms, the line-shape function $w$ must replace the Dirac
delta-function in Eq.~(\ref{eqIarpes6})---so as to phenomenologically
account for interactions---where the quantity $\delta \varepsilon_i$
denotes the characteristic width of the broadened spectrum.  Finally,
the label ``liq'' in Eq.~(\ref{wliq}) refers to the fact that we are
describing the liquid-like portion of the response of the electron
assembly. Here we will use the Lorentzian as the line shape:
\begin{equation} \label{Lorentz} w_\text{liq}(x) = \frac{\Gamma/\pi}{x^2 + \Gamma^2}
\end{equation}
thus neglecting the frequency dependence, if any, of the imaginary
part of the electron's self energy. In this work, we set $\Gamma =
0.6$~eV. This corresponds to $l_\text{mfp} \approx 9$\AA~ by
Eq.~(\ref{de}) near $k_F$.

After putting together the above notions, we obtain for the
photocurrent due to the Fermi-liquid component of the overall
electronic response:
\begin{equation}
\label{Itotaldeloc}
\begin{split}
I_\text{liq} (E, \hbar\omega_\sph) &= 2 \frac{V}{v_c} \, \frac{k_{f}
  (q_e A_0)^2}{m} (\bk_f \bm e)^2 \left| \widetilde
\psi_\text{fr}(\bk_f) \right|^2   \\ & \times w_\text{liq}(E - \varepsilon_f(\bm k_f) -
\Phi), \delta \varepsilon_i) \theta((\Phi + E_F + \hbar\omega_\sph)-E)
\times \theta(E - (\Phi + \hbar\omega_\sph))
\end{split}
\end{equation}
where the value of $\bk_f$ is fixed by energy conservation per
Eq.~(\ref{enconservout}), combined with the constraint that
$\bk_f-\bk_i = \bm G$, and by the location of the detector.  A
summation over all possible values of the reciprocal lattice vector
$\bm G$ is implied.  In practice, there will be ordinarily just one
such vector, if any. The multiplicative factor of 2 in the front has
appeared because we have explicitly summed over the electron spin.

Finally, the step functions $\theta$ are included in
Eq.~(\ref{Itotaldeloc}) to explicitly enforce that the effective
energy of the initial state be contained within the occupied portion
of the pertinent band. (We place the bottom of the band at
$\varepsilon=0$, per Eq.~(\ref{disp}).) Note the first step function
should be replaced by the Fermi distribution at finite
temperatures. In any event, because $\delta \varepsilon_i > 0$, the
line-shape envelope $w_\text{liq}$ will generally extend outside the
occupied portion of the band, even if centered on an occupied
orbital. Thus the overall contribution of the electron liquid to the
photocurrent---as integrated over the {\em occupied} orbitals---will
be less than when $\delta \varepsilon_i = 0$. The decrease will be the
more significant, the closer the nominal peak's maximum
$\varepsilon_i(\bk_i)$ is to the edges of the occupied part of the
band and, in particular, to the Fermi energy.  Conversely, some
photocurrent will be observed that corresponds to wings of spectral
lines centered at nominally empty orbitals, $\varepsilon_i(\bk_i) >
E_F$ and $\varepsilon_i(\bk_i) < 0$.

\subsection{Photocurrent due to localized sources, and its value
relative to the contribution of the Bloch states}

When the initial state is localized in space, one must compute the
transition matrix element for an individual initial state and then sum
over the latter states to obtain the total photocurrent due to this
mechanism.  Using the general expression (\ref{eqLoc4}), one readily
obtains that the photocurrent due to an individual localized orbital
can be expressed through the Fourier transform of the orbital at
$\bk=\bk_f$:
\begin{equation}
\label{M2loc}
|M_{f, i}|^2 = \left(\frac{q_e A_0\hbar}{m}\right)^2 (\bk_f \bm e)^2
\left| \: \int d^3\br \, e^{-i \bk_f \br} \psi^\text{(loc)}_i (\bm r)
\right|^2
\end{equation}
and we have indicated that the summation over the initial states is
now a discrete sum over distinct localized sources of photocurrent.

As discussed in the main text, the effective energy of a localized
state is distributed around the Fermi energy: $\varepsilon_i \approx
E_F$. The broadening is due to the states being short-lived and due to
the recoil of the Fermi-liquid. We denote this distribution with
function $w_\text{sol}(x)$:
\begin{equation} \int dx \: w_\text{sol}(x) = 1,
\end{equation}
Here we make the simplest assumption of the spectral shape above being
uncorrelated from the location in space or detailed shape of the
respective wave function. Thus the solid-like contribution of the
photocurrent is given by the expression
\begin{equation}
\label{Isol}
I_\text{sol} (E, \hbar\omega_\sph) = \frac{V}{v_c} \, \frac{k_{f} (q_e
  A_0)^2}{m} n_l \: (\bk_f \bm e)^2 \la \left| \widetilde
\psi_\text{loc}(\bk_f) \right|^2 \ra w_\text{sol}(E - (\Phi + E_F +
\hbar\omega_\sph))
\end{equation}
where $n_l$ is the number of localized sources per cell and the
averaging is over distinct shapes of the effective wavefunction
$\psi_\text{loc}(\bm r)$. Note that values of the momentum of the
outgoing electron for the delocalized and localized sources of
photocurrent are generally different, at the same value of the photon
frequency. The difference is, however, not very large because $\hbar
\omega_\sph \gg E_F$.

In contrast with the Fermi liquid case, the spectral broadening
encoded by the function $w_\text{sol}$ in Eq.~(\ref{Isol}) does not
lead to a decrease in the total amount of the corresponding
photocurrent since, by construction, we consider occupied states
only. Incidentally, the broadening due to the recoil currents is
exclusively toward the occupied side of the Fermi energy,
$\varepsilon_i \le E_F$, in the first place. We use Eq.~(2.7) from
Mahan's article,~\cite{MahanShifts} in which we set $g = 0.5$ and
$\epsilon_0 = 0.85$~eV. This form exhibits the requisite low-energy
asymptotics while being integrable. Additionally, we convolute this
form with a Gaussian with standard deviation $0.15$~eV to
phenomenologically account for the spectral-diffusion broadening,
discussed in the main text, as well as the finite resolution of the
detector.

To compare the contribution of the localized sources of photocurrent
to that of the extended ones, we adopt a specific functional form for
the wavefunction of the initial states. As explained in the main text,
an appropriate functional form is represented by the Gaussian
function:
\begin{equation} \label{psiloc}
  \psi_\text{loc}(\bm r) = \psi_q(\br) \equiv (q^2/\pi)^{3/4} \,
  e^{-q^2 r^2/2}
\end{equation}
and so,
\begin{equation} \label{aver} \la \left| \widetilde
\psi_\text{loc}(\bk_f) \right|^2 \ra = \la \frac{8 \pi^{3/2}}{q^3}
e^{-k_f^2/q^2} \ra_q,
\end{equation}
while the averaging is now with respect to the parameter $q$.  The
expression in the angular brackets effectively describes the
dependence of the amount of recoil from the localized charge proper on
the degree of localization. Judging from the sigmodal dependence of
the exponential on the r.h.s, we see the recoil is dominated by
configurations of the electron fluid such that $q \gtrsim k_f$.  We
quantify this large-$q$ response by performing the averaging above
using the probability density of the frontier orbital, in the momentum
space, as discussed in the main text. This probability density is
simply the magnitude squared of the Fourier transform of the frontier
wavefunction, normalized to unity. Given the definition
(\ref{FTphifr}), the normalization factor is $(2 \pi)^{-3}$ yielding
for the effective probability distribution
\begin{equation} \label{pq}
  p(q) = 4 \pi q^2 \left|\widetilde {\psi}_\text{fr} (q)\right|^2/(2 \pi)^3.
\end{equation}
and, consequently:
\begin{equation} \label{aver2} \la \left| \widetilde
\psi_\text{loc}(\bk_f) \right|^2 \ra = \frac{4}{\pi^{1/2}}
\int_0^\infty \frac{dq}{q} e^{-k_f^2/q^2} \, \left|\widetilde
    {\psi}_\text{fr} (q)\right|^2
\end{equation}
Finally, we obtain for the photocurrent:
\begin{equation}
  \label{Isol2}
  \begin{split}
I_\text{sol} (E, \hbar\omega_\sph) &= \frac{4}{\pi^{1/2}} \:
\frac{V}{v_c} \frac{k_{f} (q_e A_0)^2}{m} \: (\bk_f \bm e)^2
\int_0^\infty \frac{dq}{q} e^{-k_f^2/q^2} \, \left|\widetilde
    {\psi}_\text{fr} (q)\right|^2 \\ & \times w_\text{sol}(E - (\Phi + E_F +
    \hbar\omega_\sph))
  \end{split}
\end{equation}
where the value of $\bk_f$ is fixed by energy conservation:
\begin{equation} \varepsilon_f(\bm k_f) - E_F = \hbar\omega_\sph
\end{equation}
and we adopt the primitive cell as our repeat unit; thus $n_l =1$.

Aside from the aforementioned complications due to line broadening
accompanying the liquid response, the ratio of the cumulative
intensities of the liquid-like and solid-like responses, respectively,
reduces to a rather simple expression, according to
Eqs.~(\ref{Itotaldeloc}) and (\ref{Isol2}):
\begin{equation} \label{liqsolratio1}
  \frac{I_\text{liq}}{I_\text{sol}} = \frac{\pi^{1/2}}{2} \:
  \frac{k^3_{f, \text{liq}}}{k^3_{f, \text{sol}}} \frac{\left|\widetilde
        {\psi}_\text{fr} (k_{f, \text{liq}})\right|^2}{ \int_0^\infty
        \frac{dq}{q} e^{-k_{f, \text{sol}}^2/q^2} \, \left|\widetilde
             {\psi}_\text{fr} (q)\right|^2}
\end{equation}
This expression is given as Eq.~(\ref{liqsolratio}) of the main text.
We see the solid response is an integral measure of the electronic
wavefunction and, thus, is less sensitive to its detailed form than
its liquid counterpart.  The actual contributions of the liquid- and
solid-like types of response will also depend on the relative amount
of time the electron spends in the classically forbidden and allowed
regions, respectively, as discussed in the main text. In addition, the
details of line broadening will also affect the appearance of the
spectra.  While we expect the liquid-like response to yield an
approximately Lorentzian shape, the contribution of the photocurrent
due to the localized states has a more complicated functional
form. The respective peak will be relatively narrow near the maximum
but will also have a much broader base, because of the power-law
spectrum of the recoil currents. This will act to make the peak's
maximum more prominent.

\section{Plasma oscillations as a symmetry-lowering perturbation
  of the Fermi-liquid state}
\label{fermi}

Plasma oscillations are considerably faster than the motions of
valence electrons, and so, by construction, we regard the
quasi-equilibrium, Fermi-liquid response of the electron fluid as
effectively averaged over all instantaneous configurations of the
plasmons. Furthermore, on these long time scales, the electronic
excitations can be thought of as relatively well defined
quasiparticles that are decoupled from the plasmons and whose mutual
interaction is screened owing to the plasma
oscillations.~\cite{PinesEXS, nozieres1999theory}

On timescales of photoemission, however, plasmons are largely static
objects that lower the symmetry of the molecular field.  One may
inquire how the Fermi-liquid part of the electronic response is
affected by this circumstance. An upper bound for the effects of the
aforementioned symmetry lowering can be obtained by assuming the
electrons can instantly adjust to the plasmon configuration. Under
these circumstances, the plasmon simply plays the role of an
externally imposed static potential, analogously to how in the
Born-Oppenheimer approximation, we regard the field due to nuclei as
static. This construct can be also usefully looked at from the
viewpoint of the density-functional theory. Suppose we can solve the
electronic problem with an added source field. We can then perform a
Legendre transform so as to obtain the energy of the system as a
function of the spatial profile of the electron density.~\cite{GHL}

\begin{figure}[t]
  \centering
  \includegraphics[width=0.8\textwidth]{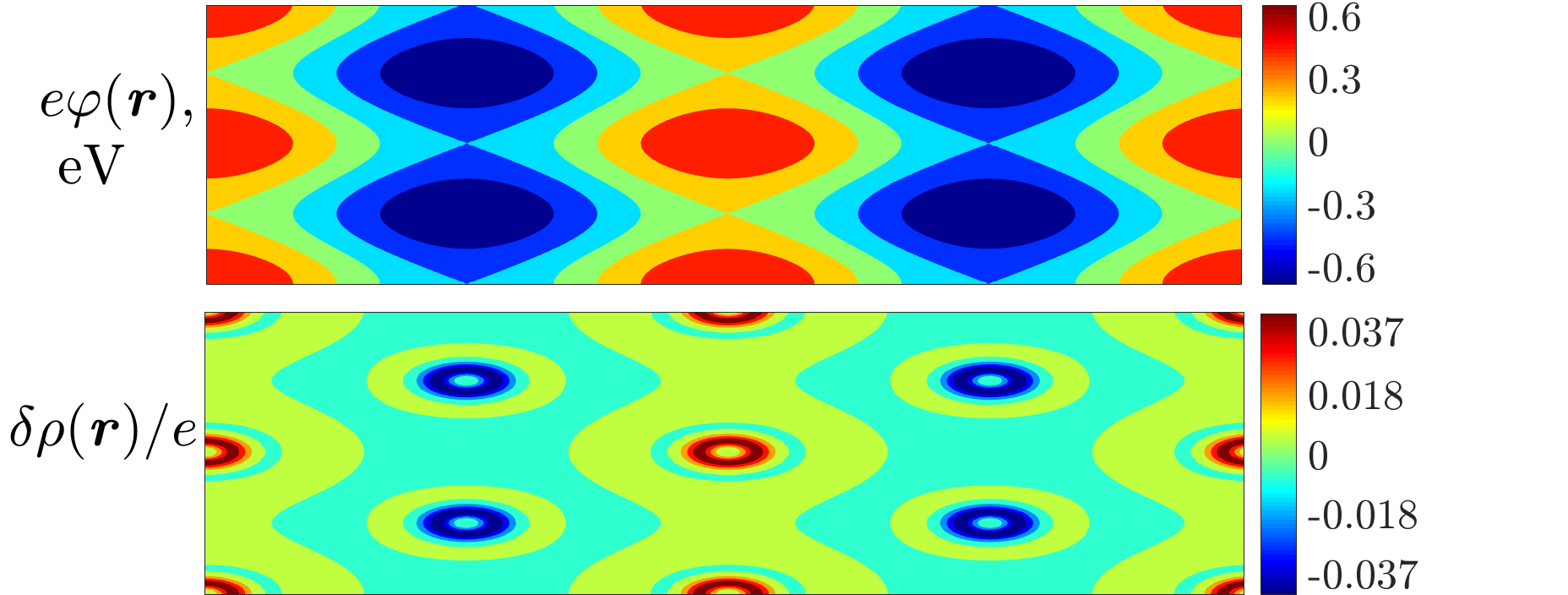}
  \caption{Examples of externally applied field $\varphi(\br)$ and the
    resulting deviation $\delta \rho$ of charge density from its
    $\varphi=0$ value in potassium, slice along (1, 1, 0). The
    magnitude of the potential 1.29~eV corresponds to the onset of the
    Van Hove singularity.}
  \label{onsiteCDW}
\end{figure}

Here we specifically consider a perturbation that lowers the symmetry
of the BCC structure found in Na and K, to that of the CsCl
structure. The latter can be thought of as two inter-penetrating cubic
lattices, one made of element A and the other of element B, where each
corner of the A lattice is in the center of the cubic cell of lattice
B and vice versa. We use the package CP2K~\cite{CP2KMainReference} to
implement this construct. Fig.~\ref{onsiteCDW}(a) shows the spatial
profile of the perturbing potential, while the
Fig.~\ref{onsiteCDW}(b) shows the resulting change in the electron
density.

\begin{figure}[t]
  \centering \includegraphics[width=0.7\textwidth]{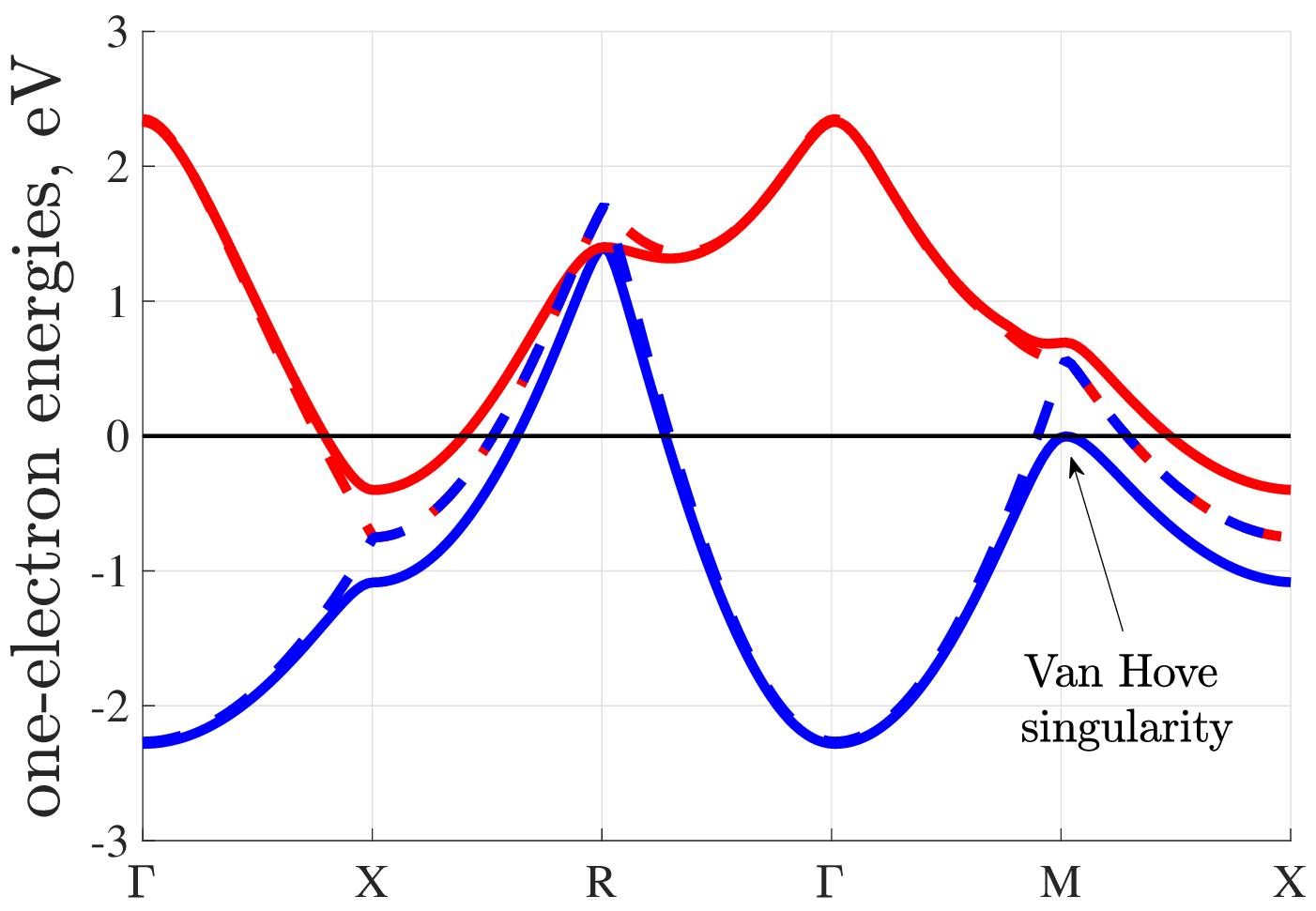}
  \caption{The one-electron spectrum of elemental K with and without
    externally imposed potential shown, respectively, by the solid and
    dashed lines. The magnitude of the potential is $1.29$~eV.}
  \label{VanHove1}
\end{figure}

One can directly observe that the perturbation illustrated in
Fig.~\ref{onsiteCDW} leads to a distortion of the Fermi surface. For
fields greater than a certain threshold value, the Fermi surface
intersects the boundary of the Brillouin zone and is no longer simply
connected while there appears a Van Hove singularity, see
Fig.~\ref{VanHove1}. If present in a static spectrum, such a
singularity would give rise to a sharp feature for a vertical
transition starting from the stationary point on the spectrum. If off
resonance, the feature would still persist---if spectral line is
broadened---but its integrated intensity would be scaled down by
magnitude of the line-shape profile at the energy corresponding to the
Van Hove singularity. In addition, the singularity is tied to the edge
of the Brillouin zone (of the CsCl structure) in terms of the momentum
but not in terms of energy, which will further smear contributions of
peak due to individual charge-density configurations.

To quantify this upper bound on these non-adiabatic effects on the
Fermi-liquid, one needs to average over all configurations of the
fluid, such as that in Fig.~\ref{onsiteCDW}(b), subject to the
aforementioned density functional. We have performed such a
calculation using a particular realization of the molecular field of
the correct symmetry~\cite{DmitrievThesis} and the specific symmetry
of the perturbing field from Fig.~\ref{onsiteCDW}.  We have found that
the overall effect of the Van Hove singularities is modest overall and
contributes to the background in the spectra but does not engender
sharp features. This finding is consistent with the notion, stated in
the main text, that the anomalous Fermi peak is a non-perturbative
effect.


\end{document}